\documentclass[]{spie}  

\usepackage{amsmath,amsfonts,amssymb}
\usepackage{graphicx}
\usepackage[colorlinks=true, allcolors=blue]{hyperref}

\usepackage{multirow}

\newcommand{\athena}{{\it Athena}}

\newcommand{\chandra}{{\it Chandra}}

\newcommand{\suzaku}{{\it Suzaku}}
\newcommand{\swift}{{\it Swift}}

\newcommand{\xmm}{{\it XMM}}

\title{The Athena WFI Science Products Module}

\author[a]{David N. Burrows}
\author[b]{Steven Allen}
\author[c]{Marshall Bautz}
\author[d]{Esra Bulbul}
\author[a]{Julia Erdley}
\author[a]{Abraham D. Falcone}
\author[b]{Stanislav Fort}
\author[c]{Catherine E. Grant}
\author[b]{Sven Herrmann}
\author[a]{Jamie Kennea}
\author[e]{Robert Klar}
\author[d]{Ralph Kraft}
\author[b]{Adam Mantz}
\author[c]{Eric D. Miller}
\author[d]{Paul Nulsen}
\author[e]{Steve Persyn}
\author[a]{Pragati Pradhan}
\author[b]{Dan Wilkins}
\affil[a]{The Pennsylvania State University (PSU), University Park, PA, USA}
\affil[b]{Stanford University, Stanford, CA, USA}
\affil[c]{MIT Kavli Institute for Astrophysics \& Space Research, Cambridge, MA, USA}
\affil[d]{Smithsonian Astrophysical Observatory (SAO), Cambridge, MA, USA}
\affil[e]{Southwest Research Institute (SwRI), San Antonio, TX, USA}

\authorinfo{Further author information: 
E-mail: burrows@astro.psu.edu \\} 

\date{}

\begin{document}	   

\maketitle

\begin{abstract}
The Science Products Module (SPM), a US contribution to the Athena Wide Field Imager, is a highly capable secondary CPU that performs special processing on the science data stream.  The SPM will have access to both accepted X-ray events and those that were rejected by the on-board event recognition processing.  It will include two software modules.  The Transient Analysis Module will perform on-board processing of the science images to identify and characterize variability of the prime target and/or detection of serendipitous transient X-ray sources in the field of view.  The Background Analysis Module will perform more sophisticated flagging of potential background events as well as improved background characterization, making use of data that are not telemetered to the ground, to provide improved background maps and spectra.  We present the preliminary design of the SPM hardware as well as a brief overview of the software algorithms under development.
\end{abstract}


\section{The WFI Science Products Module} 

The only currently approved major X-ray observatory (in the class of \xmm\ and \chandra) for the future is the ESA \athena\ mission, planned for launch near the end of the next decade.  
The \athena\ mission\cite{Nandra2018} has two instruments: an X-ray Integral Field Unit (X-IFU)\cite{Barret2018}, designed for very good energy resolution, and a Wide Field Imager (WFI)\cite{Meidinger2018}, designed for surveys requiring a very wide field of view, as well as studies of very bright X-ray sources with high time resolution. 

NASA intends to make contributions to the WFI instrument in several areas, including provision of heat pipes, design and testing of front-end ASICs for the DEPFET detectors, and a computer board called the Science Products Module (SPM), which is intended to enhance the scientific productivity of the WFI instrument by performing on-board analysis beyond what is normally done on such instruments.  Here we provide an overview of the SPM and discuss its current status.

\section{SPM Hardware}

\begin{figure}[tbp]
	    \centering
	    \parbox{6.5in}{
	    \includegraphics[width=6.4in]{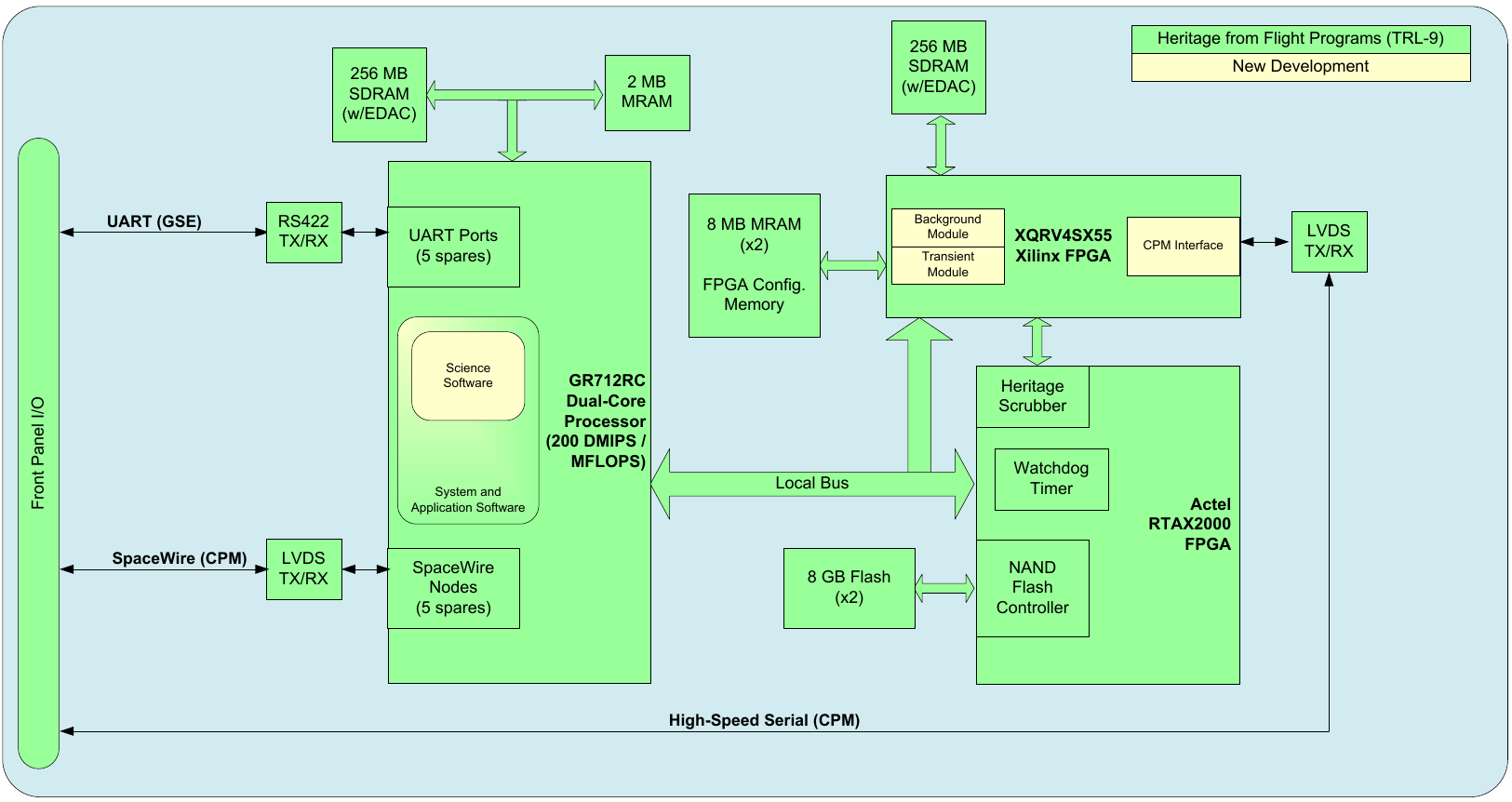}
	    \centering\parbox{6in}{
	    \caption[SPM block diagram.]
		{\small Block diagram of the Science Products Module board.  Some details (such as connections between this board and the rest of the instrument) have not been finalized.  Heritage components are shown in green, with new developments in beige.  The primary new developments are the software and firmware components of the science analysis modules described below.}
	    \label{fig:SPM_block_diagram}}} 
	\end{figure}

The SPM is based on the {\it Centaur} single board computer produced by the Southwest Research Institute (SwRI), with flight heritage from the NASA Magnetospheric Multi-Scale (MMS) and Juno missions.  
A block diagram is shown in Fig.~\ref{fig:SPM_block_diagram}.
The board includes communications, command and control, and image processing capabilities.  
The processor is a dual-core LEN3-FT CPU, providing 200 DMIPS and 200 MFLOPS combined computational performance while operating at full speed, with power dissipation below 2W.  The board has both Spacewire and custom high-speed LVDS interfaces to the main instrument controller, the Central Processing Module (CPM), allowing high-speed pixel data transfers from the Frame Processing Module in the camera electronics.  The board will be a double Eurocard format ($200 \rm{mm} \times 200 \rm{mm}$).

SwRI is responsible for designing and fabricating the SPM hardware and for providing glue firmware, boot code, system software, and command and telemetry software.  Science analysis software will be developed by Penn State, MIT, and SAO.
	
The primary purpose of the SPM is to perform on-board science analysis that can enhance the capability of the instrument.  Two analysis modules are under development: a Transient Analysis Module (TAM) and a Background Analysis Module (BAM).  These are described in the following sections.

\section{Transient Analysis Module}
\label{sec:TAM}

The Transient Analysis Module (TAM) is being developed at Penn State.  The module has two purposes: 
\begin{enumerate}
\item Monitoring the primary target for variability.  Highly variable sources will be flagged in the data stream.  There is a possibility that the detector readout mode could be changed in the case of sources that are piled up, but this is not currently the baseline.
\item Identification and preliminary characterization of transient (or highly variable) sources in the rest of the field of view.  Most of these are likely to be serendipitous sources.  We will search for the following classes of transient/variable objects:
	\begin{enumerate}
	\item Objects that vary by more than $5\sigma$ from the median count rate during the observation.  This includes objects that turn on or turn off during the observation.
	\item Objects that are significantly brighter than the catalog flux for their location on the sky.  Our baseline plan is to use the {\it eROSITA} source catalog.
	\end{enumerate} 
\end{enumerate}

\subsection{TAM processing}
\begin{figure}[bp]
	    \centering
	    \parbox{6.5in}{
	    \includegraphics[width=5in]{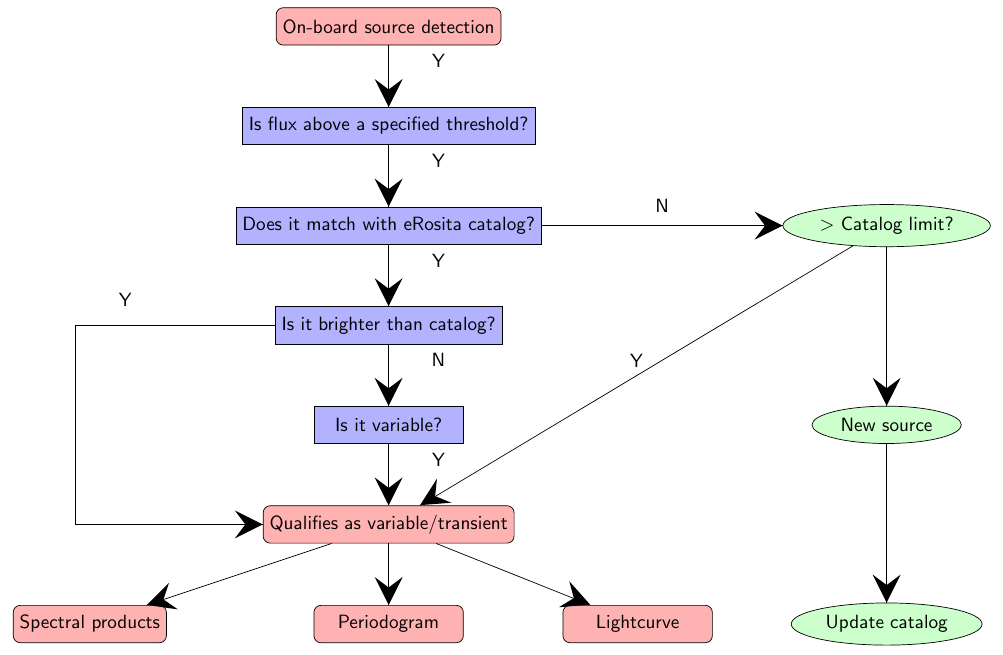}
	    \centering\parbox{6in}{\vspace{0.1in}
	    \caption[SPM flow chart.]
		{\small Top-level preliminary flow chart for the Transient Analysis Module (TAM) software.  We search for variable or transient sources; for sources that are identified as either transient or variable we produce quick-look data products for transmission to the ground.}
	    \label{fig:TAM_fc}}} 
	\end{figure}

The preliminary software flow chart for the TAM is shown in Fig.~\ref{fig:TAM_fc}.

\subsection{TAM telemetry output}

We plan to produce short telemetry messages for sources identified as being variable/transient.  These messages will be similar to the messages generated for gamma-ray bursts by the Swift XRT instrument.  They will be sent to the spacecraft on a high-priority housekeeping channel.  The intent is to have these downloaded and processed on the ground much faster than the normal science data processing.  These messages will be used to generate alerts using existing transient alert systems ({\it e.g.}, GCN, AMON, or other systems current at the time of launch).  However, since neither the spacecraft vendor nor the details of the science operations center(s) have been determined yet for the \athena\ mission, the implementation details are subject to revision.

\athena\ will be at L1 or L2, making it impossible to use near-Earth rapid response telemetry systems like the TDRSS Demand Access channel used by \swift\ to telemeter alerts.  Current plans call for one 4-hour ground contact per day to download data.  We expect to immediately telemeter data from any objects discovered on-board {\it during} these ground passes, resulting in \swift-like transient response for approximately 16\% of each day.  During other parts of the orbit, the data will be significantly delayed.

\subsection{TAM detection rates}

Preliminary estimates of detection rates for transients suggest that alerts will be dominated by AGN, with occasional detections of tidal disruption events, gamma-ray bursts, and other transients.  A rough estimate of the total rate of discovered transients can be deduced from the rate of serendipitous transients found on other missions.  For example, the \swift\ XRT instrument found 30,000 variable objects in 8 years of operation covering $> 1900$ square degrees with median sensitivity of $\sim 3 \times 10^{-13}$ erg cm$^{-2}$ s$^{-1}$\cite{Evans2014}.  This amounts to an average of 10 variable objects per day.  The \athena\ WFI instrument will have nearly 100$\times$ more collecting area and 2.9$\times$ the field of view compared to the \swift\ XRT.  In practice, we may need to set on-board thresholds to send only the most interesting (or brightest) detections to the ground for distribution of alerts.
	
\newpage	
\section{Background Analysis Module}
\label{sec:BAM}

The Background Analysis Module is being developed at MIT and SAO.  Its purpose is to use data available on-board, but not telemetered to the ground, to either reduce or characterize the instrumental background.

The instrumental background is dominated by events related to high energy charged particles traversing the instrument.  Preliminary estimates of the background level, developed using GEANT4 simulations of the expected cosmic rate interactions with a model of the instrument, indicate that the background might exceed the requirement, which is  $< 5 \times 10^{-3}$ counts s$^{-1}$ cm$^{-2}$ keV$^{-1}$ in the 2 -- 7 keV band for 60\% of the observing time.  However, much of this background is associated with events that are temporally related to minimally ionizing particles (MIPs) hitting the detector.  Because of the thickness of the silicon absorber in the DEPFET detectors, these particles deposit charge well in excess of that deposited by X-rays in the instrumental bandpass, and these saturated pixels can be easily discarded on-board.  Unfortunately, these particles also tend to produce associated ``splashes'' of charge deposition that are indistinguishable from X-ray events and are therefore not discarded by the standard on-board processing (see Fig.~\ref{fig:charged_particle_track}).  Our goal is to identify more sophisticated techniques that will allow us to either reduce the background below the requirement or improve the background characterization, without reducing source counts from valid X-ray events.

\subsection{Flight background from current missions}
\begin{figure}[bp]
	    \centering
	    \parbox{6.5in}{
	    \includegraphics[width=2.5in]{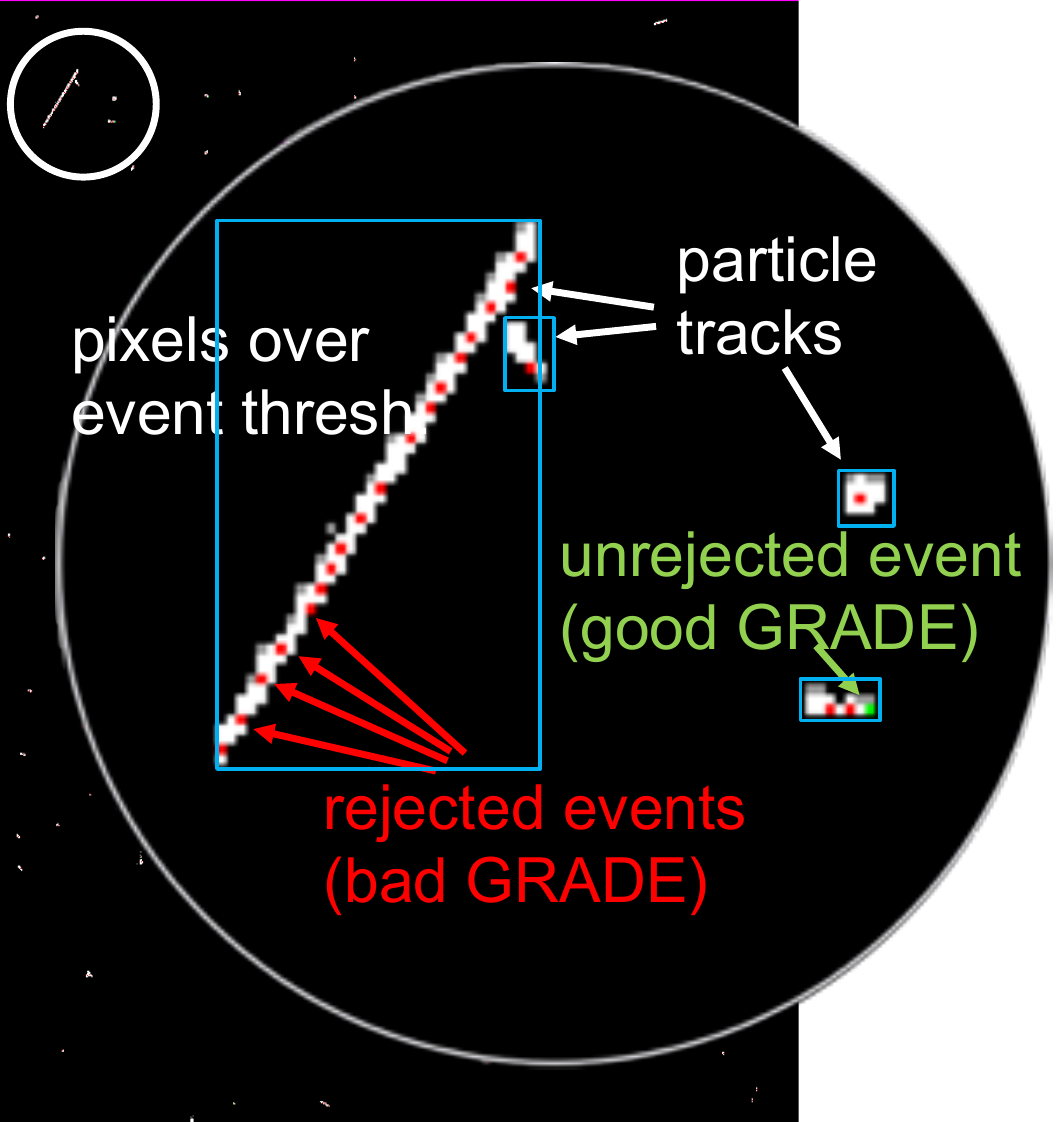}
	    \centering\parbox{6in}{\vspace{0.1in}
	    \caption[SPM flow chart.]
		{\small Example of a charged-particle track with associated pixels that are identified by the on-board event recognition algorithm as being valid X-ray events.  The large circle is zoomed in on the small circle in the upper left corner of the image.  This region contains several particle tracks (a long linear one from a particle at grazing incidence to the detector, plus several more circular events from particles incident closer to the detector normal), as well as both rejected and unrejected "events" ({\it i.e.}, pixels that are local maxima).  From Grant {\it et al.}, this volume\cite{Grant2018}.}
	    \label{fig:charged_particle_track}}} 
	\end{figure}

Preliminary investigations have focused on available data from \swift, \suzaku, 
\xmm, and \chandra, all of which send some raw, unprocessed images to the ground for calibration/instrument characterization purposes.  We have collected these data sets and examined them for charged particle events.  While the detectors involved are quite different from the WFI DEPFETs, especially in the case of the \swift, \suzaku, and \chandra\ CCDs, which have comparatively thin depletion regions and small pixels, they have the advantage of providing direct measurements of effects produced by bombardment of real detectors with real cosmic rays.\footnote{We note that the \xmm\ EPIC pn-CCD detectors are much more similar to the WFI DEPFET detectors in depletion thickness, pixel size, and frame rate.}   These efforts have focused on trying to identify spatial correlations between putative X-ray events and charged particle (MIP) tracks (Fig.~\ref{fig:charged_particle_track}). Preliminary results are discussed by two other papers in this volume\cite{Bulbul2018, Grant2018}.

\subsection{GEANT4 simulations}
The WFI team has invested a large amount of effort in developing GEANT4 simulations of cosmic ray interactions with the instrument.  While these have the disadvantage that they may not correctly simulate all of the physical interactions, they have the distinct advantage compared to use of existing data sets that they do simulate the correct detector characteristics (including depletion depth), instrument configuration (including shielding), and orbit.  We are beginning to work on applying the analysis techniques we have used on the raw images discussed above to the outputs of GEANT4 simulations.  We expect that a combination of both techniques will be used to guide development of on-board event rejection algorithms that can be applied in the SPM.

\subsection{Advanced techniques}
We are also beginning to investigate alternative approaches to this problem, including the application of neural networks.  This work is being led by Stanford.  We will explore the possibility that these more sophisticated algorithms might be better at identifying particle-related events than the simpler algorithmic solutions based on spatial correlations.  This work will use the GEANT4 simulations to train the neural net, since only in the simulations can we identify which pixels contain charge from the charged particles and which contain charge actually produced by X-rays.


\section{Conclusions}

Significant progress has been made on the SPM conceptual design over the past $\sim 18$ months, and we believe that the SPM will be able to produce interesting results in transient astrophysics and will be able to make significant improvements in the WFI background.  The next steps are to prepare inputs for the WFI Preliminary Requirements Review in September 2018, and to continue design and development work for the software analysis modules.  A review will be held in the spring of 2019 to determine whether the SPM can meet its requirements and whether it will be included in the instrument design.  This decision leads into the instrument Preliminary Design Review in June 2019.

\section*{ACKNOWLEDGEMENTS}
This work was made possible by support from NASA grant NNX17AB07G.

\bigskip
\bigskip

\bibliographystyle{spiebib} 
\bibliography{spm_references}

\begin{thebibliography}{1}

\bibitem{Nandra2018}
{Nandra}, K.~{\it et al.}., ``{Athena: observing the hot and energetic universe
  with ESA's next generation x-ray observatory},'' in [{\em Space Telescopes
  and Instrumentation 2018: Ultraviolet to Gamma
  Ray}{\nolinebreak\hspace{0.1em}]},  {\em \procspie} {\bf 10699},  10699--48
  (July 2018).

\bibitem{Barret2018}
{Barret}, D.~{\it et al.}., ``{The Athena x-ray integral field unit (X-IFU)},''
  in [{\em Space Telescopes and Instrumentation 2018: Ultraviolet to Gamma
  Ray}{\nolinebreak\hspace{0.1em}]},  {\em \procspie} {\bf 10699},  {10699--51}
  (July 2018).

\bibitem{Meidinger2018}
{Meidinger}, N.~{\it et al.}., ``{Development of the wide field imager
  instrument for Athena},'' in [{\em Space Telescopes and Instrumentation 2018:
  Ultraviolet to Gamma Ray}{\nolinebreak\hspace{0.1em}]},  {\em \procspie} {\bf
  10699},  10699--50 (July 2018).

\bibitem{Evans2014}
{Evans}, P.~A., {Osborne}, J.~P., {Beardmore}, A.~P., {Page}, K.~L.,
  {Willingale}, R., {Mountford}, C.~J., {Pagani}, C., {Burrows}, D.~N.,
  {Kennea}, J.~A., {Perri}, M., {Tagliaferri}, G., and {Gehrels}, N., ``{1SXPS:
  A Deep Swift X-Ray Telescope Point Source Catalog with Light Curves and
  Spectra},'' {\em \apjs}~{\bf 210},  8 (Jan. 2014).

\bibitem{Bulbul2018}
{Bulbul}, E.~{\it et al.}., ``{Characterizing particle background of ATHENA WFI
  for the science products module: Swift XRT and XMM-PN full frame
  observations},'' in [{\em Space Telescopes and Instrumentation 2018:
  Ultraviolet to Gamma Ray}{\nolinebreak\hspace{0.1em}]},  {\em \procspie} {\bf
  10699},  10699--157 (July 2018).

\bibitem{Grant2018}
{Grant}, C. E.~{\it et al.}., ``{Reducing the ATHENA WFI background with the
  science products module: lessons from Chandra ACIS and Suzaku XIS},'' in
  [{\em Space Telescopes and Instrumentation 2018: Ultraviolet to Gamma
  Ray}{\nolinebreak\hspace{0.1em}]},  {\em \procspie} {\bf 10699},  10699--158
  (July 2018).

\end{thebibliography}

\end{document}